\documentclass[reprint,superscriptaddress,amsmath,amssymb,aps,nofootinbib,prx]{revtex4-1}
\usepackage{amsmath}
\usepackage{amsthm}
\usepackage[colorlinks=true,urlcolor=blue,citecolor=blue,linkcolor=blue]{hyperref}
\usepackage[shortlabels]{enumitem}
\usepackage{amsfonts}
\usepackage{graphicx}
\usepackage[caption = false]{subfig}
\usepackage{multirow}
\usepackage{float}
\usepackage{bbold}
\usepackage{stmaryrd}
\usepackage{color}
\usepackage{hyperref}
\usepackage{verbatim}
\usepackage{cases}
\usepackage{amsfonts}
\usepackage{amssymb}
\usepackage[]{mathrsfs}
\usepackage[table]{xcolor}
\usepackage{epsfig}
\usepackage{bm}
\usepackage{setspace}
\usepackage{enumerate}
\usepackage{dcolumn}
\usepackage{bm}
\usepackage{enumitem}
\usepackage{multirow}
\usepackage{multirow, array} 
\usepackage{graphicx}
\usepackage{dcolumn}
\usepackage{bm}
\usepackage{color}
\usepackage{amssymb}
\usepackage{lipsum}
\usepackage{amsmath}
\usepackage{graphicx}
\usepackage{array}
\usepackage{multirow}
\usepackage{booktabs}

\usepackage[english]{babel}
\begin{document}

\preprint{APS/123-QED}

\title{Complete numerical description of the laser-induced thermal profile in a liquid, to explain complex self-induced diffraction patterns}

\author{J. L. Dom\'{i}nguez-Ju\'{a}rez}%
\affiliation{Centro de F\'isica Aplicada y Tecnolog\'ia Avanzada, Universidad Nacional Aut\'onoma de M\'exico, Boulevard Juriquilla 3001, Juriquilla, 76230 Quer\'etaro, M\'exico}

\author{M. A. Quiroz-Ju\'{a}rez}%
\affiliation{Centro de F\'isica Aplicada y Tecnolog\'ia Avanzada, Universidad Nacional Aut\'onoma de M\'exico, Boulevard Juriquilla 3001, Juriquilla, 76230 Quer\'etaro, M\'exico}

\author{J. L. Arag\'{o}n}%
\affiliation{Centro de F\'isica Aplicada y Tecnolog\'ia Avanzada, Universidad Nacional Aut\'onoma de M\'exico, Boulevard Juriquilla 3001, Juriquilla, 76230 Quer\'etaro, M\'exico}

\author{R. Quintero-Berm\'udez}%
\affiliation{Department of Chemistry and Physics, University of California, Berkeley, California 94720, USA}

\author{R. Quintero-Torres}%
\email{rquintero@fata.unam.mx}
\affiliation{Centro de F\'isica Aplicada y Tecnolog\'ia Avanzada, Universidad Nacional Aut\'onoma de M\'exico, Boulevard Juriquilla 3001, Juriquilla, 76230 Quer\'etaro, M\'exico}


\date{May 11, 2022}

\begin{abstract}
In the past, laser propagation in a fluid with heat transfer has been modeled using simplistic conduction and convection conditions yielding inaccurate predictions. Here we present a detailed numerical study describing the thermal profile of the fluid and its interaction with the laser. Furthermore, we evaluate the diffraction field in the far field produced by a pump beam impinging in the fluid and the interferometric pattern obtained normal to the propagation direction to test our model. Direct comparison between experimental results and numerical simulation allows for a complete understanding of the energy transfer from the laser to the liquid and the subsequent effect on the laser propagation. Spatial self-phase modulation and propagation control from small-phase diffraction to aberration-controlled diffraction and up to diffraction oscillation are observed and explained with our modeling.
\end{abstract}

\maketitle



\section{Introduction} 
The complex nature of natural convection phenomena in closed spaces has been extensively discussed \cite{bergman2011introduction, tien}. The complexity arises due to the coupling of the thermal variables with the flow variables, independent boundary conditions of thermal and flow phenomena, and the strong effect of small changes in the initial conditions \cite{hernandez2015natural, proskurnin2015advances}. The effect of light propagation on thermal transport further increases the complexity of the problem due to a variable optical path length that the beam itself generates. All this makes both the theoretical and experimental investigation of the spatial self-phase modulation in the propagation of the laser beam in a liquid extremely difficult and, at the same time, very rich.
There is a qualitative understanding of the phenomena associated with heat transport in a liquid thanks to multiple experiments and analytical studies that describe the various elements of the entire process \cite{ostrach1972natural, pretot2000theoretical, kozanoglu2007thermal}. A remarkable work was due to Edward Lorenz \cite{lorenz1963deterministic}. In his original work, Lorenz modeled the properties of a two-dimensional fluid cell of uniform depth warmed from below and cooled from above. Specifically, this mathematical model is a reduced version of Barry Saltzman's system \cite{saltzman1962finite}, whose equations can be derived from the Navier-Stokes equation for fluid flow by considering the Oberbeck–Boussinesq approximation and severe truncations of Fourier series for the hydrodynamic fields \cite{hilborn2000chaos}. Despite its importance, this model only provides a simplified description of heat transfer phenomena under very special conditions, which prevents gaining a deep understanding of the involved processes in order to export them to other problems. At this respect, a detailed description of light propagation in a liquid that considers heat transport is lacking. A detailed description of the propagation of a laser beam in a liquid requires knowledge of the variable optical path length (or refractive index) that fully explains the physics behind diffraction, diffraction variation, and diffraction oscillations. Here we present a complete image of the interaction of a pump laser with the liquid to explain these phenomena.
Laser absorption in a medium has been associated as the cause of the diffraction effect, discounting that in most cases absorption is only the beginning of a cascade of energy transfer processes that transforms the excitation into a thermal profile that introduces a phase in the propagation $(x)$ of the laser $\phi(y,z)=k\int_{x_1}^{x_2} n(x,y,z) dx$. This introduced phase added to the initial laser wavefront defines the diffraction profile \cite{deng2005formation, ara2012diffraction}.
Absorption can be intrinsic as in a homogeneous material or extrinsic as is the case with dyes or nanomaterials \cite{xiao2021near}, but the thermal change in materials is controlled by both the size of the laser interaction and the cell dimensions in addition to the photothermal properties of the sample. Nonlinear optical phenomena can have other contributions than thermal, but it is accepted that thermal is the largest and, in turn the slowest \cite{akhmanov1968self}.

\section{Methods}

\subsection{Experimental setup }
\begin{figure}[t]
\includegraphics[width=8cm]{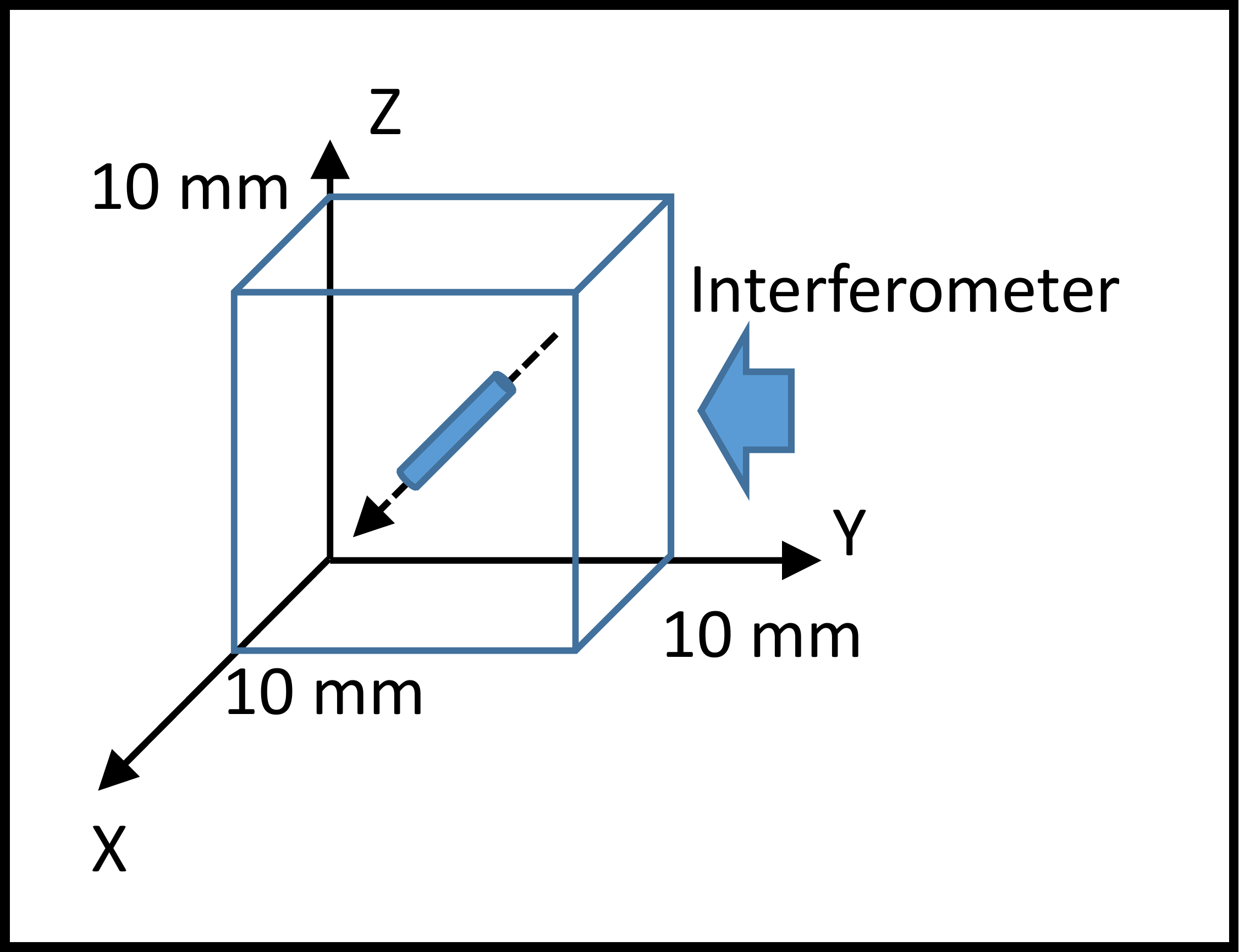}
\caption{The cube represents the cell-sample frame. The cylinder represents the pump laser parallel to X-axis, and the Michelson interferometer has one arm parallel to Y-axis.}
\label{fig:figure1}
\end{figure}

A standard spectroscopic cuvette holds a liquid solution, water or ethanol. Ethanol has a larger refractive index change with temperature; however, pure water will present a similar effect with a proper adjustment in the pump power and absorption coefficient. We define the pump laser propagation direction as the x-axis, the z-axis as that aligned with the height of the solution and it is opposite to the gravity direction, where $z = 0$ indicates the bottom of the cell. The y-axis is aligned with one of the interferometer arms and $y = 5$ mm indicates the plane where the laser is propagating, as is presented in figure \ref{fig:figure1}. Experiments are conducted with IR medium intensity (40 mW) for heating and supercontinuum for diffraction, as well as 520 nm laser for the Michelson interferometer.

The refractive index sensitivity of ethanol at 25 $^o$C is $S=dn/dT=$-3.9X10$^{-4}$ [1/K] \cite{bialkowski2019photothermal}. A sudden transition from OFF to ON of a pump laser with a cross-section of 60 $\mu$m, will produce a shift of one whole fringe $(2\pi)$ in a Michelson interferometer. This shift indicates a change in the temperature of the ethanol due to the pump laser of $\Delta T=\lambda/2Sd^{\dagger}=-2$ [K]. $d^{\dagger}$ is about five times the laser cross section due to thermal conduction and will change with time from $d$ to $5d$.

\begin{figure}[t]
\includegraphics[width=9cm]{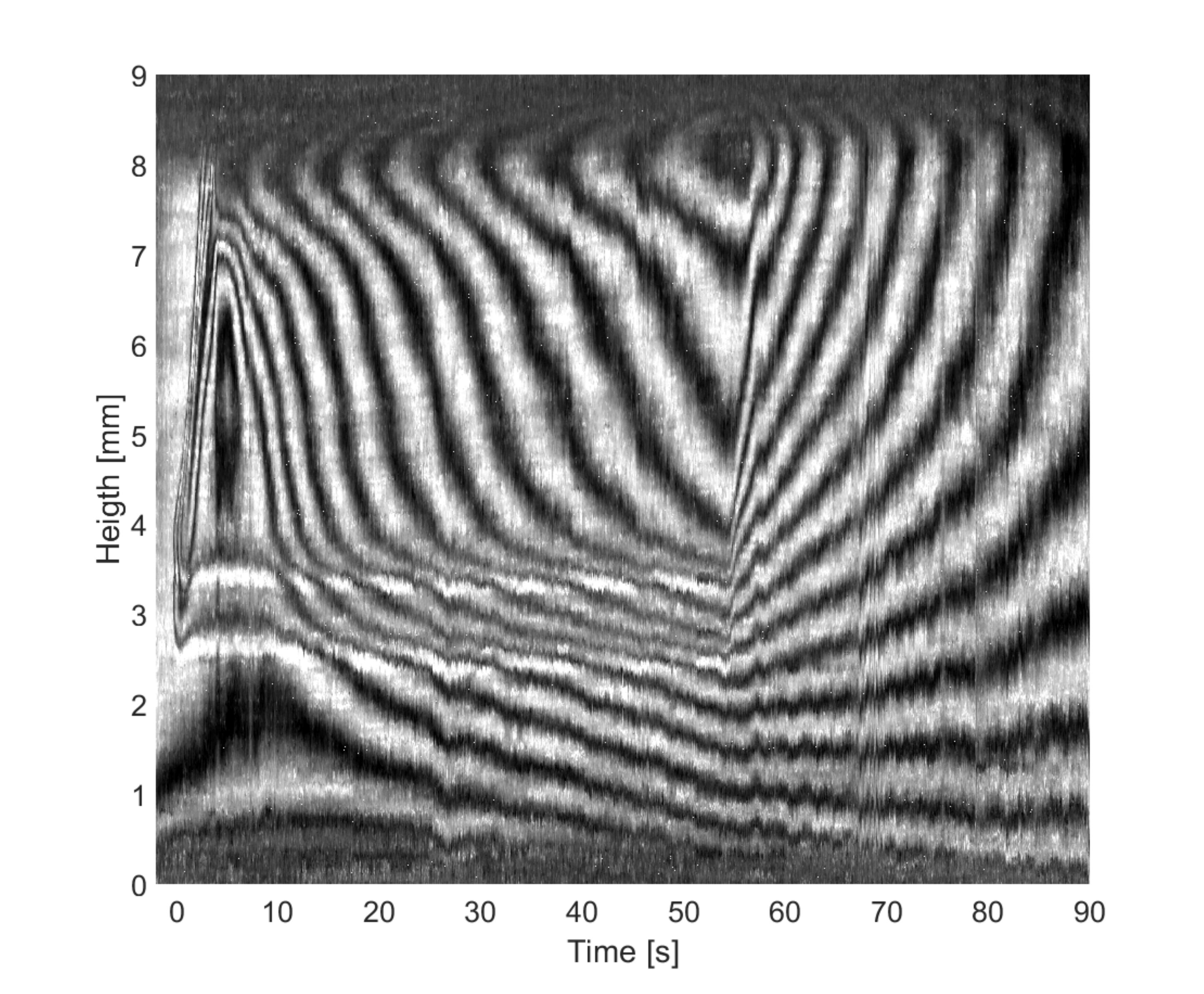}
\caption{Isophase lines produced with a Michelson interferometer. The horizontal axis represents the time, and the laser-ON ranges from 0 to 55 s. The vertical axis is the height. The link between phase and temperature is the refractive index sensitivity $S=dn/dT$, accepted as constant in a small temperature range.}
\label{fig:figure2}
\end{figure}

\begin{figure}[t]
\includegraphics[width=9cm]{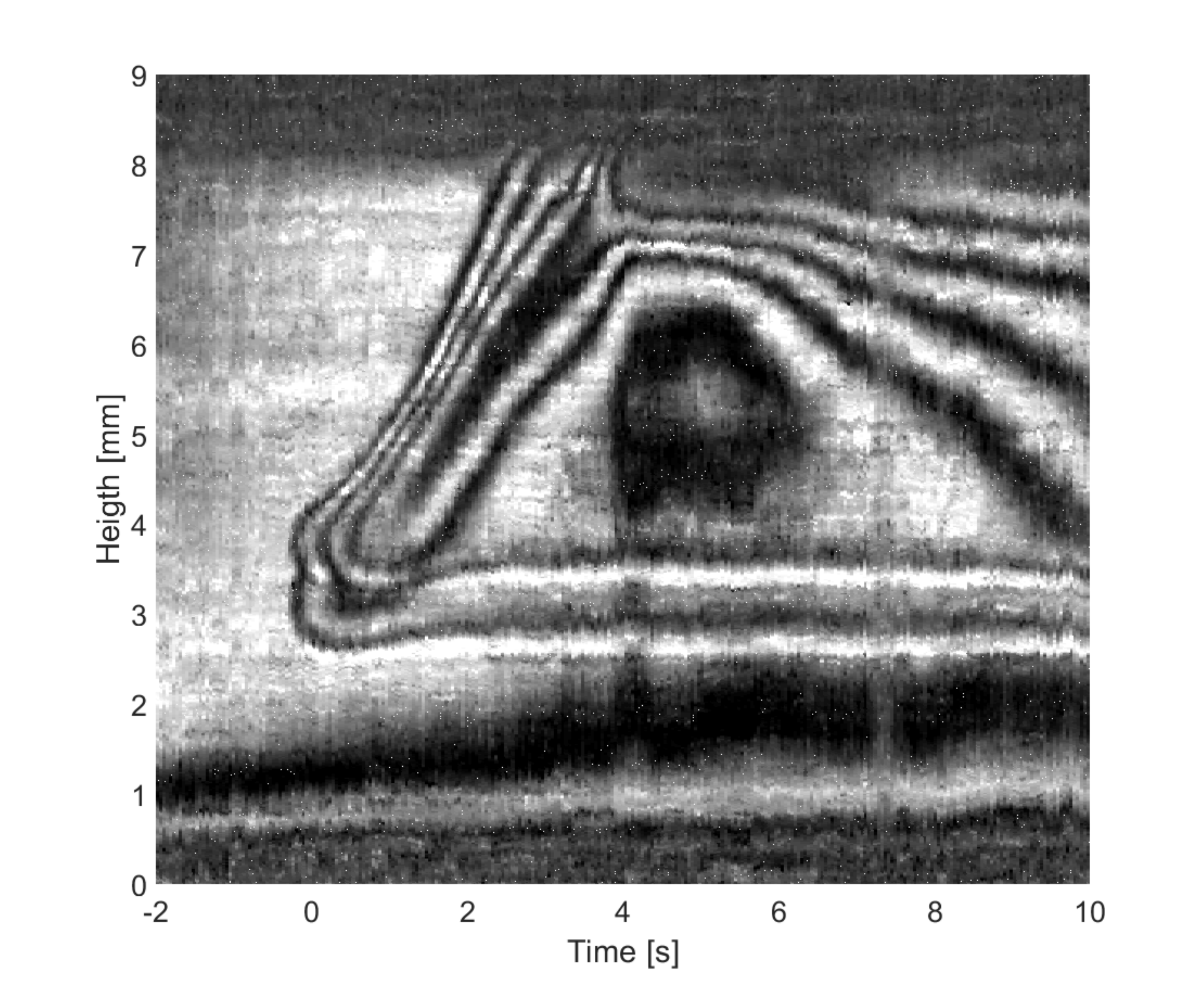}
\caption{Zoom of figure \ref{fig:figure2}, this profile illustrates the natural convection and the lack of conduction in this time region.}
\label{fig:figure3}
\end{figure}

On the other hand, at a steady state, a phase shift of 15 fringes in the Michelson interferometer is explained if the Michelson interferograms are collected during an ON-OFF pump-laser sequence at 24 fps: for two seconds before switching the laser ON, for 55 seconds with the laser ON, and for 35 seconds after the laser is switched OFF. In each interferogram, the horizontal axis is along the pump laser and the vertical axis along the solvent height ($X Z$ plane). From these images, the dynamic variation of the phase from the Michelson interferometer can be ascertained by selecting a line of information at a fixed horizontal point ($x=5$mm) and plotting its evolution with time leading to images like the one illustrated in figure \ref{fig:figure2}. Figure \ref{fig:figure3} is an expanded section in figure \ref{fig:figure2} to appreciate the ON/OFF transition.

The following observation sets the condition in which the power transferred to the cell produces diffraction and self-induced aberration. For a typical fluid (refractive index decreases with temperature, $\Delta \phi<0$) and a divergent wavefront (i.e. focus of the beam is before the sample), the diffraction image starts unchanging the incident beam. The diffraction increases monotonically by increasing the cylindrically symmetrical diffraction rings. It distorts in the direction of gravity by decreasing the radius, then it further distorts the difraction as in aberrations produced by a tilted lens. An oscillating diffraction pattern can be produced in any of the above situations depending on the distance between the beam and the meniscus, requiring less power to generate oscillations as this distance decreases \cite{gouesbet2001instabilities}.

\subsection{The model}

A standard heat transfer description is applied. The equations for the fluid that include conservation of mass, linear momentum, and angular momentum are \cite{lukaszewicz2016navier}:

\begin{eqnarray}
\frac{d\rho}{dt}+\nabla  \cdot (\rho \vec{u})&=&0, \label{eq:rho}\\
\rho \frac{d\vec{u}}{dt}+\rho\vec{u}\cdot\nabla\vec{u}&=&\nabla\cdot\sigma+F_V+\rho g, \\
\sigma^T&=&\sigma,
\end{eqnarray}

where the variables are taken from the typical notation; $\rho$ [kg/m$^3$], $\vec{u}$ [m/s], $\sigma$[N/m$^2$], $F_V$ [N/m$^3$], $g$ [m/s$^2$], respectively represent density, velocity vector, Cauchy stress tensor, the external volumetric force and the gravity acceleration.

\begin{figure}[t]
\includegraphics[width=9cm]{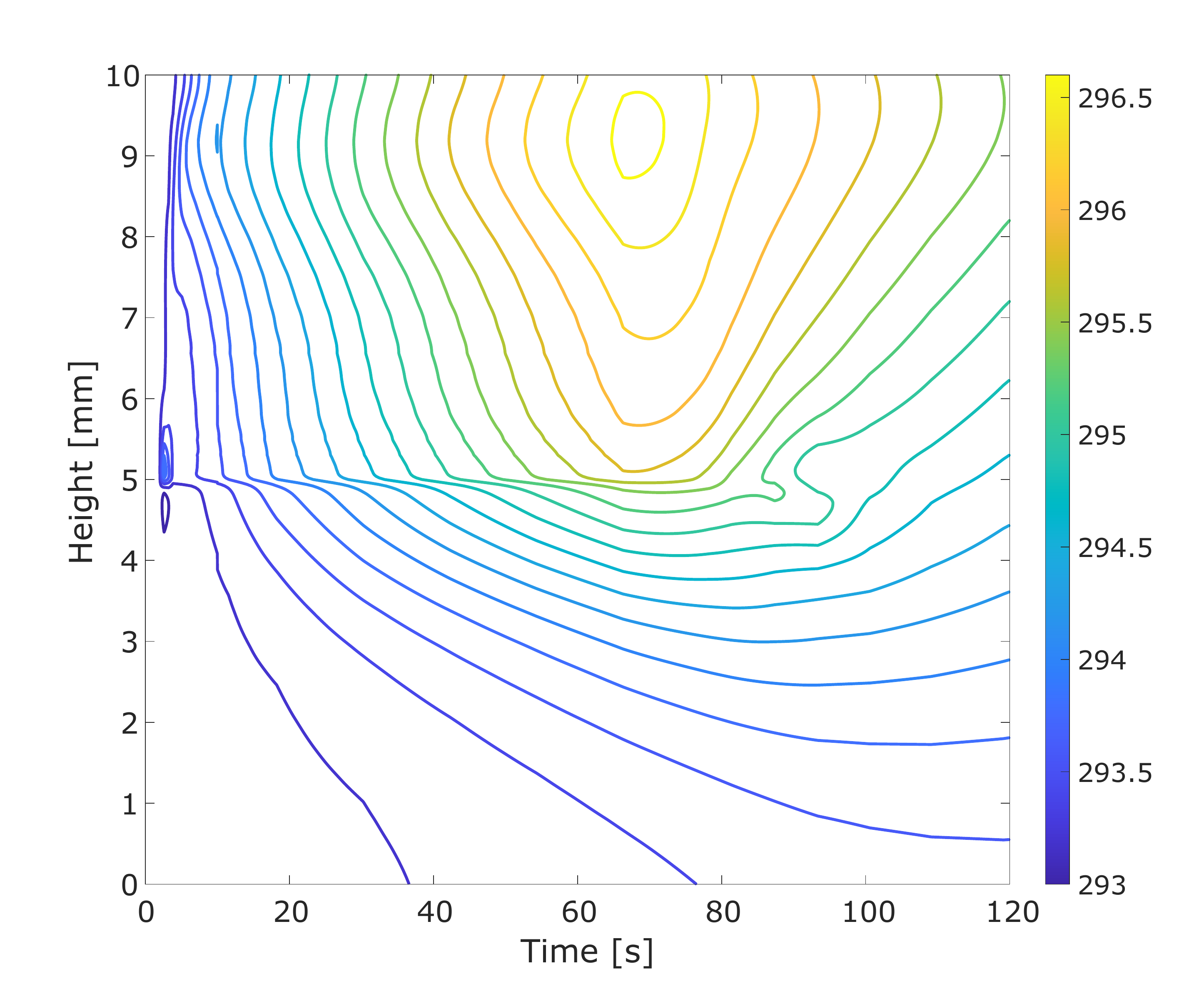}
\caption{Simulated temperature profile in a setup resembling figure \ref{fig:figure2}, if this result is accepted, many new details are revealed as essential to understand the phase change in the pump laser. The isothermal step is 0.2 K.}
\label{fig:figure4}
\end{figure}

The equation for the thermal variables is:

\begin{equation}
    \rho C_p \frac{dT}{dt}+\rho C_p \vec{u}\cdot \nabla T=\nabla \cdot (k\nabla T)+ \dot{Q}_V
    \label{eq:temperatura}
\end{equation}

where the variables use the standard notation for $C_p$ [J/kgK], $T$ [K], $k$ [W/mK], $\dot{Q}_V$ [W/m$^3$] as mass heat capacity, temperature, thermal conductivity, and the volumetric power, respectively.

To simplify the problem, all the interfaces are modeled by means of the heat transfer coefficient, $h$ [W/m$^2$K], where $\dot{Q}_A$ [W/m$^2$] is the power crossing a surface: 

\begin{equation}
    \dot{Q}_A=h(T_{ext}-T)
\end{equation}

\begin{figure}[t]
\includegraphics[width=9cm]{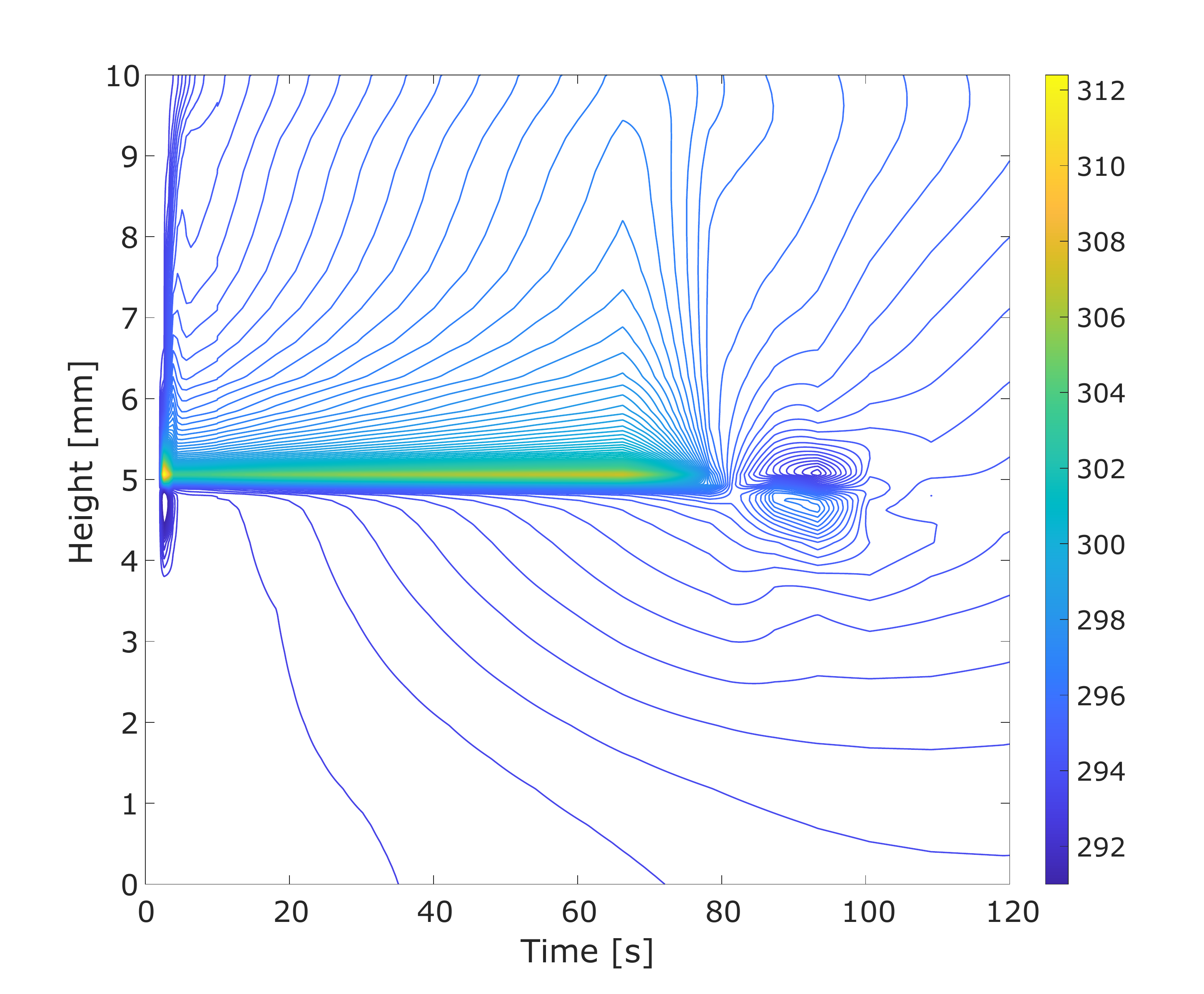}
\caption{Simulated temperature profile on the $X$-$Z$ plane with $Y=5$ mm. This is the real source of change in the propagation of the pump beam. The initial changes last about 5 seconds, and the drift with time remains in a small degree. The isothermal step is 0.2 K.}
\label{fig:figure5}
\end{figure}

\subsection{Numerical simulation}

To solve Equations (\ref{eq:rho}) to (\ref{eq:temperatura}), we used COMSOL Multiphysics software \cite{multiphysics3}. So, the numerical simulation couples the heat transfer equation and the Navier Stokes equations to examine the thermal profile of the fluid. The numerical simulation was performed with heat transfer coefficient $h=50$ [W/(K m$^2$)] at every interface except the liquid-air interface that uses $h=5$ [W/(K m$^2$)]. The laser beam was described by the volumetric power:

\begin{equation}
    \dot{Q}_V=4x10^9\left(\frac{x_R^2}{x_R^2+\left(x-x_0\right)^2}\right)\left(e^{-10x}\right)\left(t<60\right)
\end{equation}

where $x_0$ fixes the focal position of the beam, $x_R=\pi \omega_0^2n/\lambda$ is the Rayleigh range and the beam waist radius is $\omega_0=30$ $\mu$m.

\section{Results}
The numerical simulation of the equations can be used to determine the temperature, the velocity vector and the pressure, provided that the temperature dependencies of density, heat capacity, thermal conductivity and dynamic viscosity are known. In this case, there were 25,000 spatially dispersed points with a temporal resolution of 100 fps. The small temperature gradient was far from turbulence, so laminar flow was preferably used.

\begin{figure}[t]
\includegraphics[width=9cm]{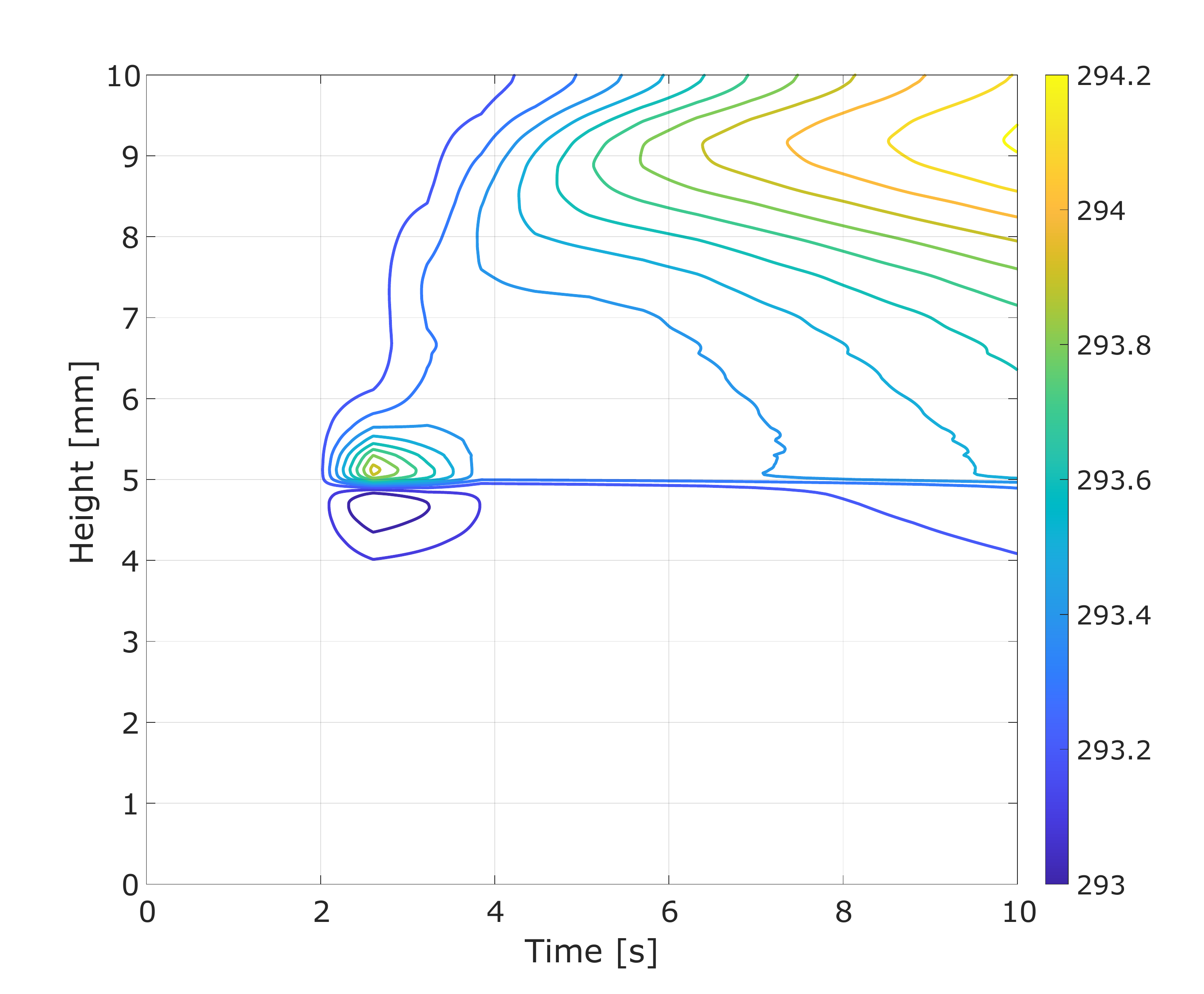}
\caption{Simulated temperature profile on the cell. This simulation was done at 100 fps, twice as fast as the camera used. The steady change with time does not indicate the pump laser propagation. The isothermal step is 0.1 K.}
\label{fig:figure6}
\end{figure}

Figure \ref{fig:figure4} depicts the average temperature along y for each value of $z$ as a function of time obtained from the simulation results, $T(z,t)$. The similarity to the experimental results (Figure \ref{fig:figure2}) can be readily observed. Figure \ref{fig:figure6} is a magnification of the first 10 seconds of the simulation, this figure explains the peculiar interferogram in figure \ref{fig:figure3} at the start of the laser incidence, the phase produced over (hot) and under the laser (cold) due to the temperature details.

\begin{figure}[t]
\includegraphics[width=9cm]{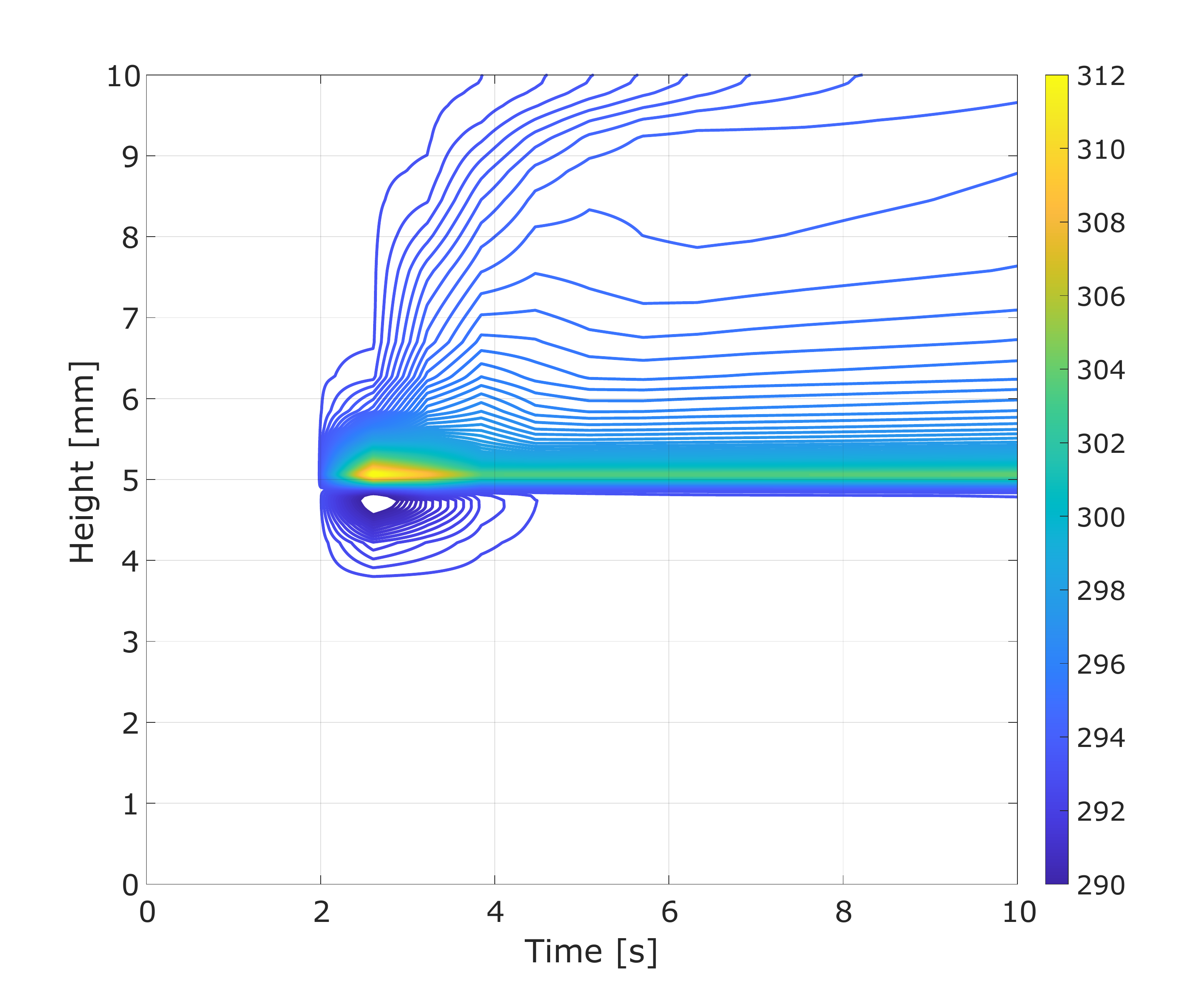}
\caption{Simulated temperature profile on the X-Z plane with Y=5 mm. This is the real source of change on the pump beam. The initial changes from the 2 to 4 seconds should be considered but subsequent variation is small. The isothermal step is 0.2 K.}
\label{fig:figure7}
\end{figure}

Figure \ref{fig:figure5} shows the temperature in the plane of laser propagation $T(x=0.5,y=0.5,z)$, this is the path that modifies the phase of the laser and produces diffraction when considering the volume through which the laser passes. Figure \ref{fig:figure7} shows the magnification of the first 10 seconds of figure \ref{fig:figure5}. Again, the asymmetry in relation to the direction of gravity can be readily observed.

Figure \ref{fig:figure8} helps to explain why the simulation matches the experiment, it shows the temperature details near the beam propagation at 10 s. The cylindrical frame represents the cross-section of the beam with a radius of about 30 $\mu$m, while on the horizontal scale, the minor division is 200 $\mu$m. The thermal maximum is almost 80 $\mu$m above the optical center. This description is valid for the before mentioned deposited power, the thermal maximum and the center of the beam coincide at low power and the thermal profile extends beyond the cross-section of the beam. On the other hand, symmetry with respect to y holds at all powers. likewise, in this figure, the slanted temperature profile can be used to justify the aberrations of the beam in which they resemble an ideal lens inclined with respect to the propagation of the laser.

\section{Discussion}
The pump beam modifies its properties when passing through the cell, if the power transferred to the cell is minuscule, the beam maintains its properties and passes without modification; when the power transferred to the cell is moderate, the transmitted beam expresses symmetrically cylindrical diffraction; only changing the intensity distribution as a Fresnel aperture or as an opaque barrier depending on the position of the laser focus relative to the center of the cell. That is, a positive curvature of the wavefront, as is the case of a divergent beam in a medium whose refractive index increases with temperature, produces diffraction similar to a Fresnel aperture, with identical behavior if both factors change sign. In the case of factors with different signs, diffraction occurs as in an opaque barrier.

By increasing the power transfer to the cell, the diffraction pattern is distorted, resembling aberrations such as that of a tilted lens, the borders resemble a funnel, by further increasing the power transfer and with a small distance between the beam and the meniscus, oscillations of the diffraction pattern are produced.

\begin{figure}[t]
\includegraphics[width=9cm]{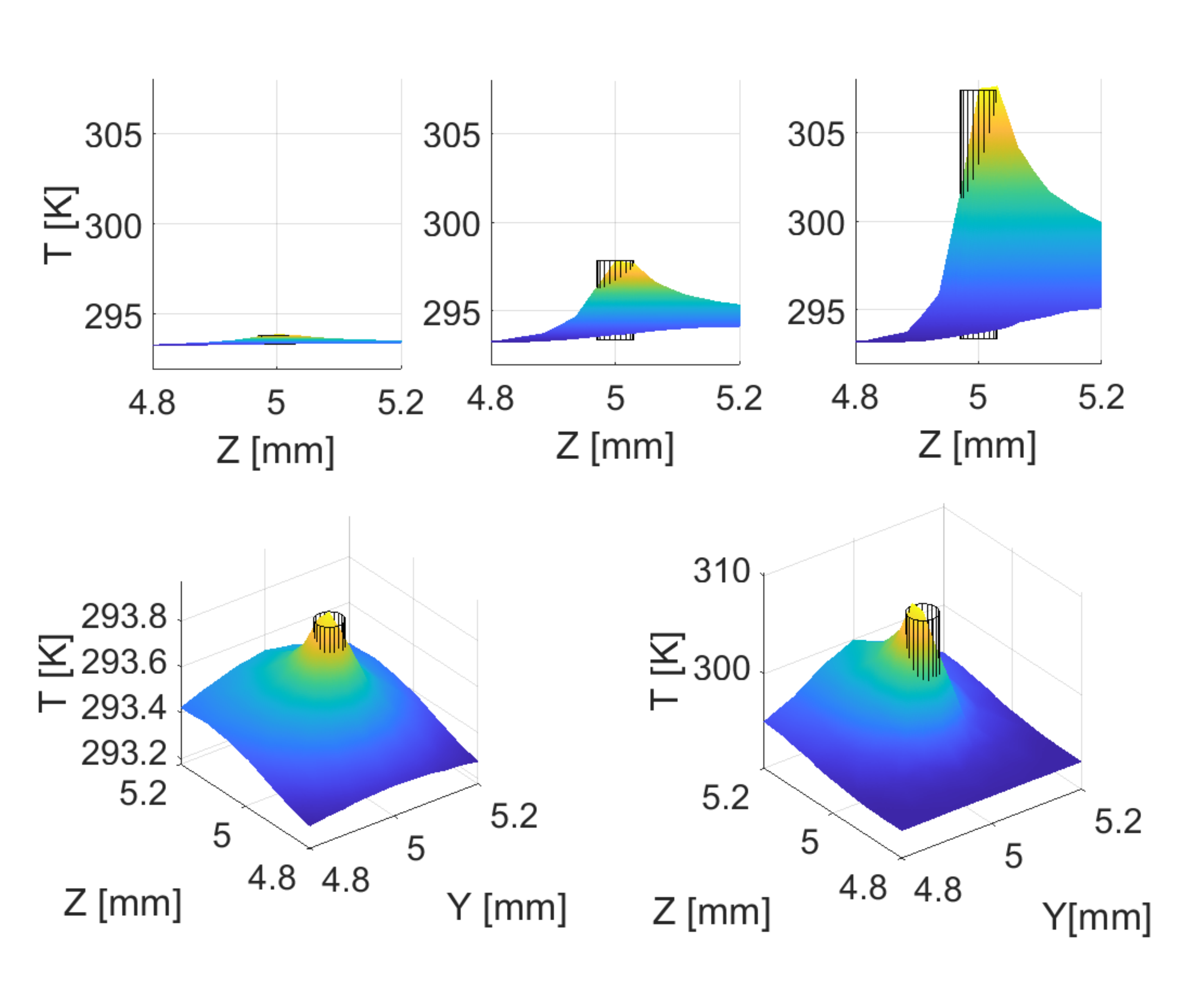}
\caption{Simulated temperature profile on the $Y$-$Z$ plane, with $X=5$ mm. The top three are for 10 s and constant in the volumetric power of 1X10$^8$, 1X10$^9$, and 4X10$^9$, the cylinder has a 30 $\mu$m radius and the asymmetry and astigmatism increase with power. The bottom two are the same as the top one and the tree.}
\label{fig:figure8}
\end{figure}

The diffraction of the pump beam is due to spatial self-modulation of the phase, which is controlled by the dynamics of the thermal profile in the cell, so understanding the origin and distribution of the profile is essential. Optical methods reveal the contribution of all changes accumulated along the path. As a result, the Michelson interferogram reveals the changes along the entire traversed distance to the laser. In contrast, the changes in the pump laser are only defined by the changes in the region that is illuminated. The main problem in the description of diffraction is the lack of details of the thermal profile along the path of the pump beam.

Given the complex nature of the equations that control the thermal profile, numerical simulations are essential to understanding the origin of the diffraction details. The experimental results shown in Figures \ref{fig:figure2} and \ref{fig:figure3} and the similarity with the simulation presented in Figures \ref{fig:figure4} to \ref{fig:figure8} confirm that they reveal the unknown details of the thermal profile and that they help explain the diffraction patterns.

The thermal profile is symmetric with respect to the y-coordinate, the time variable is decisive for the reproducibility since the thermal changes are very important in the beginning, the effects of natural convection, viscosity and conduction are clearly visible in figure \ref{fig:figure2}. Figure \ref{fig:figure3} illustrates that, relative to the laser axis, natural convection is seen at the top and conduction is seen at the bottom. The simulation allows visualizing the thermal profile of only the laser axis. In addition to the dynamics of the thermal profile, the asymmetry in the z coordinate is seen when moderate power is transferred to the cell.

It is important to insist that the laser allows the phenomenon to be visualized indirectly since the effect in the pump beam and the interferometer is produced by an extended path.

\section{Conclusions}

Comparing the experimental results and the numerical simulation, it is possible to clearly understand the results of the optical probes that reveal the thermal changes of the cell. With this information it is possible to explain the diffraction patterns that are produced when a beam transfers power to a liquid, this dynamically modifies its thermal profile. In turn this thermal profile modifies the propagation of the incident beam.
Assuming zero phase shift, no matter the curvature of the wavefront, the beam propagates linearly. If the phase change follows a Gaussian profile such as that defined by a slight change in temperature, the diffraction pattern formed will be defined by this and the curvature of the wavefront. If the temperature change is larger, the phase change is defined by the sum of the refractive indices along the beam path in this profile and stops being cylindrically symmetric. Larger power transfer to the sample and small distance to the meniscus produce periodic variations in the phase that are translated into periodic patterns in the diffraction index.

\section*{Acknowledgment}
J.L.D.J. thanks Catedras CONACYT-UNAM. R.Q.-T and J.L.D.J. wish to thank the financial support from CONACYT, under grant A1-S-8317. M.A.Q.-J. would like to thank the support from DGAPA-UNAM under Project UNAM-PAPIIT TA101023.  



\nocite{*}

\bibliography{apssamp}

\end{document}